# Soliton blue-shift in tapered photonic crystal fiber

**S. P. Stark, A. Podlipensky and P. St.J. Russell**

Max Planck Institute for the Science of Light

Guenther-Scharowsky Str. 1/Bau 24, 91058 Erlangen, Germany

http://mpl.mpg.de

We show that solitons undergo a strong blue shift in fibers with a dispersion landscape that varies along the direction of propagation. The experiments are based on a small-core photonic crystal fiber, tapered to have a core diameter that varies continuously along its length, resulting in a zero-dispersion wavelength that moves from 731 nm to 640 nm over the transition. The central wavelength of a soliton translates over 400 nm towards shorter wavelength. This accompanied by strong emission of radiation into the UV and IR spectral region. The experimental results are confirmed by numerical simulation.

Photonic-crystal fibers (PCFs) are useful for manipulating the spectral and temporal characteristics of ultrashort pulses [1]. They have been used in many research fields, for example in biological studies and optical coherence tomography [2, 3], in quantum optics [4] and in the observation of unconventional objects like the "peregrine" soliton [5]. The success of PCFs started with the demonstration of dramatic spectral broadening for a broad range of pulse parameters [6]. The broadening depends on the pulse duration and on the pump wavelength in relation to the zero-dispersion wavelength (ZDW). In many cases optical solitons can be observed, playing a major role in the spectral shaping mechanism [7]. In perturbation-free cases, solitons can propagate undisturbed over long distances, and collide without exchanging energy

[8]. When two solitons co-propagate, bound states can form [9]. Any frequency dependence in the group velocity dispersion perturbs the propagation dynamics [10]. For example, solitons respond to the presence of a single ZDW by emitting dispersive waves (DWs). As a result, supercontinua can be tailored to extend into the UV [6]. On the other hand, the long-wavelength edge of the supercontinuum is shaped by the spectral red-shift of the solitons that occurs in Raman-active media [11]. This soliton self-frequency shift (SSFS) can be cancelled in fibers with an extra long-wavelength ZDW [12], compensating resonant radiation being continuously shed into the IR.

In this Letter, we report unexpected soliton dynamics when the dispersion landscape varies along the direction of propagation. Fibers with slowly decreasing dispersion have been successfully used for adiabatic compression of solitons [13, 14] in taper transitions with lengths ranging from a few mm [15] to several m [16]. This has been used to extend the width of a supercontinuum [16] and to generate coherent, low-noise fs-pulses in the visible [15]. Tapered fibers have also been used to generate high-power parabolic pulses [17]. Here we study experimentally and numerically the behaviour of solitons in fiber transitions with two blue-shifting ZDWs, pulses being launched in the anomalous dispersion region.

The GVD curve of the untapered PCF is shown in Fig 1a, together with a scanning-electron micrograph of its microstructure. The core diameter is 1.76 μm.

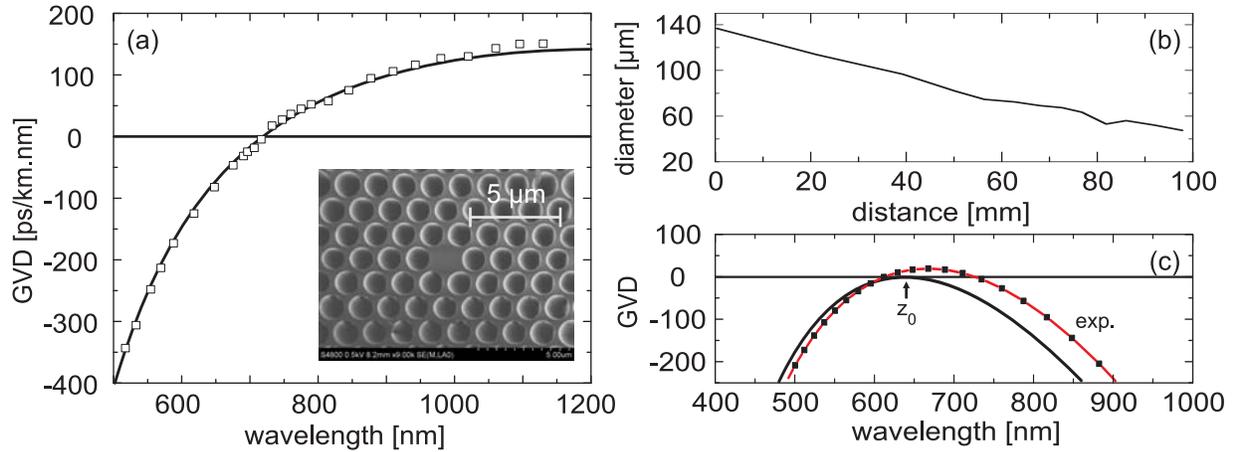

Fig. 1. (Color online) (a) GVD curve of the fiber used for the fabrication of the taper transition. The dots are the experimentally measured points, and the solid curve is a simulated dispersion profile, obtained from a silica strand model. The inset shows an SEM of the fiber. (b) Outer-fiber diameter along the transition. (c) The solid line is the simulated dispersion of the taper at its narrowest point where the two ZDWs coalesce into one at $z_0$. The dotted curve is the experimentally measured GVD curve of a fiber with a uniform 48 μm outer diameter.

The ZDW of the untapered fiber is at 735 nm. The flame-brush technique was used to fabricate a 10 cm long taper transition [18]. The profile of the fiber transition (Fig. 1b) was measured in an optical microscope. The target minimum taper waist diameter $r_{end}$ = 45 μm (core radius ~0.32 μm) was chosen to support a vanishingly narrow anomalous dispersion region, resulting from the collapse of the two ZDWs into a single point at $z_0$ (Fig. 1c). Fig. 1c shows the simulated dispersion profile at this waist diameter, calculated using a simplified model based on a circular-cylindrical strand of silica in air. Also plotted is the measured GVD for a 30 cm long uniformly tapered fiber with outer diameter 48 μm. The result is close to the desired dispersion profile at the taper end.

A mode-locked Ti:Sapphire laser system (Coherent Inc. MirA 900D), emitting pulses of duration 130 fs and energy up to 7 nJ, was used. A half-wave plate and a polarization cube were used to control the pulse energy. An additional half-wave plate was used to align the polarization state of the light along one of the principal axes of the weakly birefringent fiber. We used a 40× aspherical in-coupling lens to achieve launch efficiencies of 60%. The experimental results were simulated by numerically solving the generalized nonlinear Schrödinger equation – see [19, 20] for further details.

In Fig. 2 we show the spectral evolution along the fiber taper for two different input powers. The upper figures show the experimental cutback measurements. The white lines show the positions of the ZDWs along the taper. The taper had a ~4 cm long untapered input pigtail, which is included in the numerical simulations shown in the lower part of Fig. 2. For both power values the initial phase of pulse broadening is governed by the prevailing anomalous dispersion. In Fig. 2a the estimated experimental launched peak power is 660 W, whereas the numerical modelling (chosen for the best fit to the experimental data) is for a power of 500 W. The imperfect agreement originates from uncertainties in the values of the nonlinear coefficient ($\gamma = 0.12$ W$^{-1}$m$^{-1}$) and the launch efficiency.

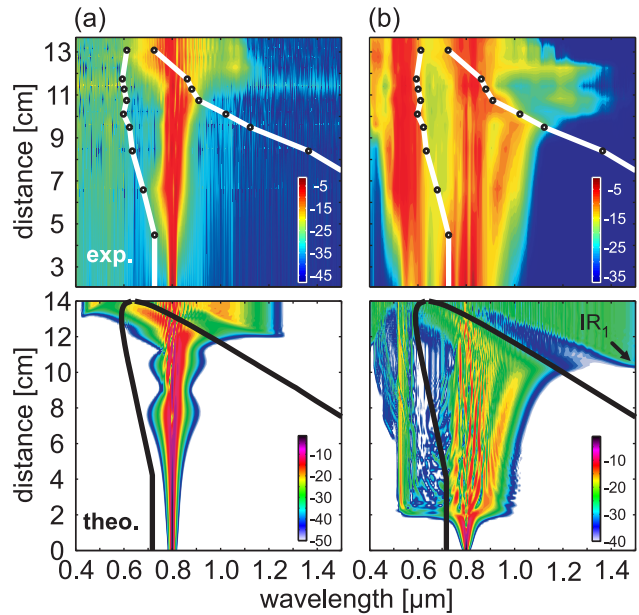

Fig.2. (Color online) Spectral evolution with distance of a 130 FWHM fs pulse launched into a taper with the profile shown in Fig. 1b. The upper plots are experimental and the lower plots numerical. The peak powers are (a) 660 W and (b) 6.5 kW. The white curves show the positions of the two ZDWs, obtained from simulations (based on SEMs of the cleaved fiber tips) using the fixed-frequency plane wave method [21]. Long wavelength bend loss causes a strong increase in IR attenuation beyond the 12 cm point.

The launched pulse (in the anomalous dispersion region) undergoes soliton fission within the taper transition at ~7.5 cm where the core radius is ~716 nm. The resulting DW wavelength is at 480 nm, calculated using the phase-matching equation [22]. Energy conversion to the DW is strongly reduced by its large detuning from the pump frequency [23], and indeed no DW can be observed.

When the power is raised to 6.5 kW (Fig. 2b), soliton fission occurs in the untapered fiber pigtail at a distance of 2.1 cm and a DW is created at 575 nm. Several fundamental solitons emerge after the fission process, the magnitude of the input power determining the number [8]. After the pulse breakup the propagation dynamics are governed by the evolution of these fundamental solitons. Each experiences the SSFS at a rate determined by the distance-dependent $\beta_2(\lambda_{sol})$ value at the soliton wavelength $\lambda_{sol}$. Note that $\beta_2(\lambda_{sol})$ changes its sign during propagation, going from positive to negative. The negative value results in an accelerated red-shift, experienced by the strongest soliton in Fig. 2b. The signal increase at the blue edge of the supercontinuum can be traced back to four-wave mixing between the soliton and the DW [22].

At a particular propagation distance the solitons sense the presence of the blue shifting ZDWs. For example, in Fig. 2a the long-wavelength part of the spectrum shows an oscillatory behaviour and acquires a "nose" pointing towards the blue. In Fig. 2b the strongest red-shifting soliton collides with the long wavelength ZDW at ~10 cm. This is accompanied by a spectral extension into the IR. Thereafter the soliton follows the blue-shift of the ZDW.

Both features are a result of the soliton undergoing a spectral blue-shift and are explained as follows. We base our discussion on the high-power results from Fig. 2b, in which the soliton blue-shift is more apparent.

In Fig. 3 shows the evolution of the centre wavelength $\lambda_{sol}$ of the two strongest solitons in Fig. 2b during propagation. They are labelled **A** and **B** and were obtained by numerically gating the solitons in the temporal domain. The SSFS causes the strongest soliton **A** to move to 1020 nm at ~10 cm (downward-pointing arrow in Fig. 3). The SSFS red-shift then halts and turns around, following the shifting ZDW, eventually reaching a wavelength of 670 nm.

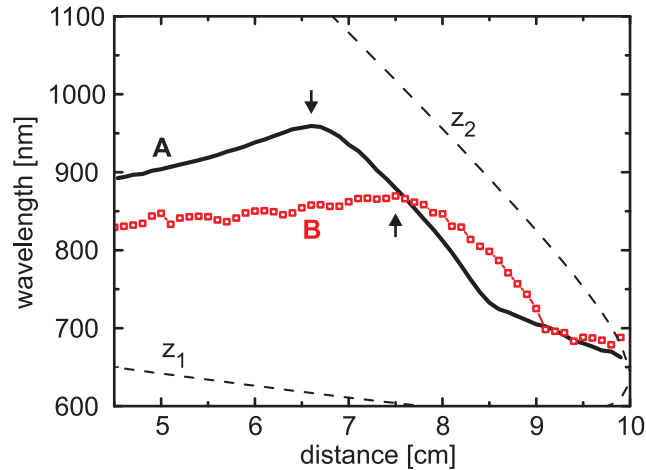

Fig. 3. (Color online) Evolution of the centre wavelength of the first two solitons (labelled **A** and **B**) emerging after the soliton fission process. After the maximum red-shift due to SSFS, a blue-shift occurs in response to the blue-shifting ZDW $z_2$.

When several solitons are present, each of them experiences a blue-shift. For example, in Fig. 3 soliton **B** moves to 920 nm at 11 cm (upward-pointing arrow in Fig. 3). After this point it moves to shorter wavelengths, eventually reaching 660 nm at 14 cm.

This blue-shift occurs at the expense of a drop in peak power, caused by the proximity of the second zero-dispersion point. As a result the soliton sheds energy continuously into the IR, producing the spectral band marked $IR_1$ in Fig. 2b. As the soliton and the long-wavelength ZDW move to shorter wavelength, the DW wavelength falls. This results in the formation of a fin-shaped band in the IR region of Fig. 2b. Note that the experimental IR signal is strongly attenuated due to long-wavelength tunnelling and bend loss [24], an effect that is not included in the numerical simulations.

The blue-shift moves the soliton into a spectral region containing the remaining pump light. The interference of the overlapping pulses causes spectral fringes, which spread out towards shorter wavelength as the core diameter falls. The blue-shifting soliton is not only responsible for the generation of light in the IR, but also leads to energy transfer into the UV. If the soliton wavelength is close to $z_1$, phase-matched generation of a DW on the blue side of the soliton is possible. Such a DW appears in Fig. 2b at 405 nm, its wavelength increasing as it propagates beyond the 12 cm point.

We now study the physics of SSFS reversal using soliton **A** (Figs. 2&3). We find that after 8 cm of propagation, soliton **A** has a FWHM duration of 17.0 fs and center wavelength of 963 nm. The core radius of the PCF is 0.693 μm at this point, with a GVD of −68.8 ps$^2$/km at 963 nm. We have simulated the propagation of this soliton in the taper transition, starting at 8 cm, in the absence of any other solitons or linear waves. The result is a clear blue-shift of the center wavelength of the pulse, accompanied by DW generation (Fig. 4a). Up to the 11 cm point, the soliton duration falls and its peak power increases. After this it loses energy due to DW generation and its peak power reduces.

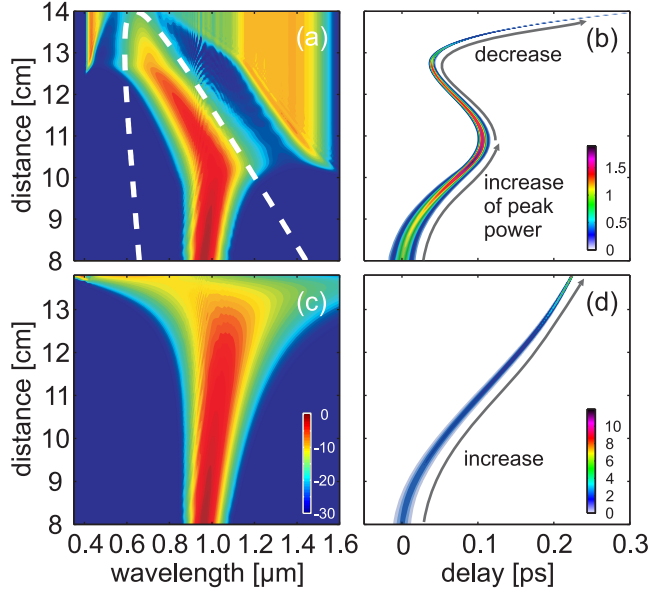

Fig.4. (Color online) (a) and (b) show the spectral and temporal evolution of soliton **A** in the 8-14 cm portion of the taper transition. In (c) and (d) the GVD profile was approximated by a frequency independent $\beta_2$, varying linearly from $-68.8$ ps$^2$/km at 8 cm to 0 ps$^2$/km at 14 cm. The nonlinear coefficient was set to a constant value in (c) and (d).

We now study the impact of the changing GVD landscape, keeping the nonlinear coefficient constant at 0.15 m$^{-1}$W$^{-1}$. Then we removed the higher-order dispersion terms, setting $\beta_2$ to a constant value of $-68.8$ ps$^2$/km. Under these circumstances there are no ZDWs and the soliton is affected neither by DW generation nor by SSFS cancellation. The transition effect is approximated by a linear variation of $\beta_2$ from $-68.8$ ps$^2$/km at 8 cm to 0 ps$^2$/km at 14 cm. The spectral and temporal evolution are shown in Figs. 4c&d. The continuous reduction in $|\beta_2|$ leads to a continuous decrease in soliton duration and an increase in spectral bandwidth. As required by energy conservation, the peak power increases (Fig. 4d). The effects of the blue and red ZDWs can be seen by comparing Figs. 4a&b and 4c&d: The soliton bandwidth is restricted, spectral components on its long wavelength edge being shed far into the IR. On the other hand, blue spectral components are continuously generated in the anomalous dispersion region, providing the seed for shifting the soliton envelope into the blue.

In conclusion, soliton-like pulses can undergo a blue-shift, opposite to the Raman-induced red-shift, when the light is launched into a fiber with a gently varying long-wavelength ZDW. The soliton wavelength can shift over a large range, accompanied by the emission of dispersive waves in the UV and IR. The mechanism is in fact a form of pulse reshaping in which freshly generated spectral components at the blue edge of the soliton cause its center wavelength to move towards the blue. The tapered fiber system could be useful for generating ultrashort solitons across the entire visible spectral range, offering fresh opportunities for tailoring light over ultrabroad spectral ranges.